\documentclass[aps,prb,reprint,
citeautoscript,superscriptaddress,showpacs]{revtex4-1}

\usepackage{graphicx,epsfig,float}
\usepackage{bm}
\usepackage{amssymb}
\usepackage{amsmath}
\usepackage[varg]{txfonts}

\allowdisplaybreaks
\renewcommand{\vec}[1]{\bm{\mathrm{#1}}}     

\begin{document}
\title{Interlayer interaction, shear vibrational mode, and tribological properties of two-dimensional bilayers with a~commensurate moir\'e pattern}

\author{Alexander S. Minkin}
\email{amink@mail.ru}
\affiliation{Keldysh Institute of Applied Mathematics of Russian Academy of Sciences,
4 Miusskaya sq., Moscow, 125047, Russia}

\author{Irina V. Lebedeva}
\email{liv\_ira@hotmail.com}
\affiliation{Simune Atomistics, Avenida de Tolosa 76, San Sebastian 20018, Spain}

\author{Andrey M. Popov}
\email{popov-isan@mail.ru}
\affiliation{Institute for Spectroscopy of Russian Academy of Sciences, Fizicheskaya str.~5, Troitsk, Moscow 108840, Russia}

\author{Sergey A. Vyrko}
\email{vyrko@bsu.by}
\affiliation{Physics Department, Belarusian State University, Nezavisimosti Ave.~4, Minsk 220030, Belarus}

\author{Nikolai A. Poklonski}
\email{Corresponding author; poklonski@bsu.by}
\affiliation{Physics Department, Belarusian State University, Nezavisimosti Ave.~4, Minsk 220030, Belarus}

\author{Yurii E. Lozovik}
\email{lozovik@isan.troitsk.ru}
\affiliation{Institute for Spectroscopy of Russian Academy of Sciences, Fizicheskaya str.~5, Troitsk, Moscow 108840, Russia}
\affiliation{Moscow Institute of Electronics and Mathematics, National Research University Higher School of Economics, Bol.~Trekhsvjatitel'skij per., 1-3/12, build.~8, Moscow, 101000, Russia}

\begin{abstract}
The potential energy surface (PES) of interlayer interaction of infinite twisted bilayer graphene is calculated for a set of commensurate moir\'e patterns using the registry-dependent Kolmogorov--Crespi empirical potential. The calculated PESs have the same shape for all considered moir\'e patterns with the unit cell size of the PES which is inversely related to the unit cell size of the moir\'e pattern. The amplitude of PES corrugations is found to decrease exponentially upon increasing the size of the moir\'e pattern unit cell. An analytical expression for such a PES including the first Fourier harmonics compatible with the symmetries of both layers is derived. It is shown that the calculated PESs can be approximated by the derived expression with the accuracy within 1\%. This means that different physical properties associated with relative in-plane motion of graphene layers are interrelated and can be expressed analytically as functions of the amplitude of PES corrugations. In this way, we obtain the shear mode frequency, shear modulus, shear strength and barrier for relative rotation of the commensurate twisted layers to a fully incommensurate state for the considered moir\'e patterns. This barrier may possibly lead to the macroscopic robust superlubricity for twisted graphene bilayer with a commensurate moir\'e pattern. The conclusions made should be valid for diverse 2D systems of twisted commensurate layers.
\end{abstract}
\maketitle

\section{Introduction}

Structural superlubricity, i.e. the mode of relative motion of the layers with vanishing or nearly vanishing friction \cite{Hirano1990, Hirano1991}, has attracted a considerable attention in the context of discovery of graphene and other 2D materials, see Ref.~\onlinecite{Hod2018} for a review. First this phenomenon was observed for nanoscale contacts between graphene flakes at the tip of a microscope probe and graphite surface \cite{Verhoeven2004, Dienwiebel2005, Filippov2008}. A wide set of atomistic simulations has been devoted to superlubricity for 2D systems with a finite size of the contact area where the edge or rim contribution to the static friction is dominant \cite{Verhoeven2004, Filippov2008, Xu2013, Koren2016, Bonelli2009, vanWijk2013, Guo2007, Shibuta2011, Wang2019, Zhang2015a, Zhang2018, Zhang2022}. Recently not only nanoscale but also micro- and macroscale superlubricity has been found in systems of 2D layers \cite{Androulidakisl2020, Liu2012, Vu2016, Song2018}.
These studies raise interest to possible factors which cause the static friction and can restrict superlubricity for a macroscale incommensurate contact area \cite{Liu2012, Mandelli2017, Hod2018, Koren2016, Minkin2021, Minkin2022}.
The following possible reasons of very low but nevertheless nonzero static friction have been considered: 1) contribution of incomplete unit cells located at the rim area of one of the layers forming a moir\'e pattern (rim contribution), \cite{Koren2016} 2) incomplete static friction force cancellation within complete unit cells of a commensurate moir\'e pattern (area contribution) \cite{Koren2016}, 3) motion of domain walls of large commensurate domains formed upon relaxation of moir\'e patterns \cite{Liu2012, Mandelli2017, Hod2018}, and 4) contribution of atomic-scale defects \cite{Liu2012, Minkin2021, Minkin2022}. The present paper is devoted to the detailed study of the area contribution to the static friction by the example of twisted graphene bilayer.

Whereas for very small twist angles, the size of the moir\'e pattern unit cell is large \cite{Campanera2007} and, therefore, formation of commensurate domains separated by incommensurate domain boundaries occurs during the structural relaxation \cite{Lebedeva2020}, for twist angles far from the coaligned orientation (0$^\circ$, 60$^\circ$ and so on), the size of the  moir\'e pattern unit cell is smaller or comparable with the width of commensurate domain walls (about 10 nm for bilayer graphene \cite{Popov2011}). For such angles, relaxation to the commensurate domains is not possible and a set of commensurate moir\'e patterns can be observed \cite{Mele2012, Campanera2007}. Twisted bilayer graphene with a commensurate moir\'e pattern has the interlayer interaction energy slightly lower than in a fully incommensurate state \cite{Xu2013}. Thus, one can expect that than such patterns can be formed preferably for the corresponding range of twist angles. Here we propose that the energetic preference of commensurate moir\'e patterns can also lead to the robust superlubricity. This is why the study of tribological properties of 2D systems with commensurate moir\'e patterns is of high interest.

The tribological properties are determined by the potential energy surface (PES) of interlayer interaction that is the dependence of this energy on coordinates describing the relative in-plane displacement of 2D layers. Previous atomistic calculations allowed to distinguish the area contribution into the static friction of twisted graphene bilayer for a few commensurate moir\'e patterns \cite{Xu2013, Koren2016, Kabengele2021}. Cancellation of the static friction force within complete unit cells of commensurate moir\'e patterns of graphene bilayer \cite{Koren2016, Minkin2021, Minkin2022} and double-walled carbon nanotubes \cite{Kolmogorov2000, Belikov2004, Bichoutskaia2006} has been also demonstrated. However, the symmetry and shape of the PES of interlayer interaction for commensurate moir\'e patterns of infinite twisted graphene bilayers has not been studied yet. Recently we proposed a hypothesis that such PESs in diverse 2D materials with layers aligned in the same or opposite directions can be universally described by the first spatial Fourier harmonics \cite{Lebedev2020}. This hypothesis has been confirmed by calculations of PESs for different 2D materials \cite{Ershova2010, Lebedeva2011, Popov2012, Lebedeva2012, Zhou2015, Reguzzoni2012, Lebedev2016, Zhou2015, Lebedev2020} and 2D heterostructures \cite{Popov13b, Jung2015, Kumar2015, Lebedev2017}. Moreover, this hypothesis is valid also for double-walled carbon nanotubes \cite{Vucovic2003, Belikov2004, Bichoutskaia2005, Bichoutskaia2009, Popov2009, Popov2012a}, where only Fourier harmonics compatible with the symmetry of both wall contribute into the PES of interwall interaction \cite{Vucovic2003, Popov2009}. By analogy with double-walled nanotubes, one might expect that PES for interlayer interaction of twisted layers of infinite commensurate moir\'e pattern is determined by Fourier harmonics compatible with symmetries of the both layers, i.e. with the symmetry of the whole moir\'e pattern. In such a case, the approximated PES depends on a single parameter and a set of physical quantities determined by the PES are interrelated \cite{Popov2012, Lebedev2016}.

Here we calculate PES for a wide set of infinite commensurate moir\'e patterns of twisted graphene bilayers using the registry-dependent Kolmogorov--Crespi potential \cite{Kolmogorov2005} and show that these PESs can be excellently approximated by the first Fourier harmonics which are compatible with the symmetry of the whole moir\'e pattern. The PES approximation derived is used to obtain analytical expressions for a set of physical quantities of moir\'e patterns determined by the PES such as the shear frequency, shear modulus, shear strength and barrier for relative rotation of the layers to a fully incommensurate state.

The paper is organized in the following way. In Sec. II,
the model of the superlubric system and calculation methods are described. Sec. III is devoted to our results on the PES calculation and approximation by Fourier harmonics as well as estimates of physical quantities determined by the PES. The conclusions and discussion are presented in Sec. IV.

\section{Methodology}

\begin{figure*}
   \centering
 \includegraphics{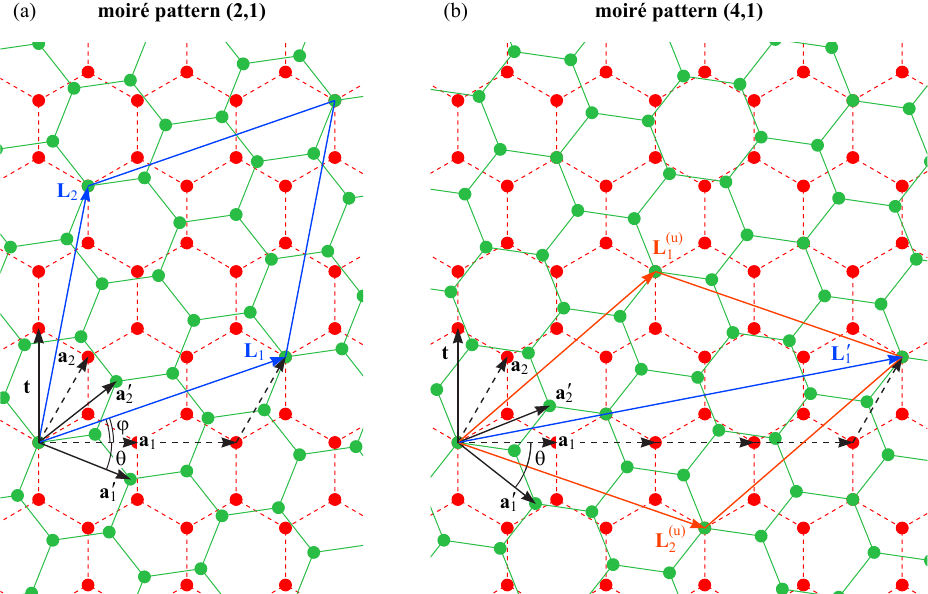}
   \caption{(Color online) (a) and (b) schemes of twin commensurate moir\'e patterns (2,1) and (4,1) of twisted graphene bilayer, respectively, with the same size of the unit cell. Lattice vectors $\vec{a}_1$ and $\vec{a}_2$ of the bottom graphene layer and $\vec{a}_1'$ and $\vec{a}_2'$ of the top layer, lattice vectors $\vec{L}_1$ and $\vec{L}_2$ of the commensurate moir\'e pattern, angle $\theta$ of relative rotation of the graphene layers and angle $\varphi$ between the lattice vector $\vec{a}_1$ of the bottom layer and lattice vector of commensurate moir\'e pattern $\vec{L}_1$ are indicated. The translational displacement $\vec{t}$ of the upper layer that converts one moir\'e pattern into the other are shown by black arrows.}
   \label{fig:01}
\end{figure*}

\subsection{Structure of commensurate twisted graphene bilayer}

Let us consider first the structure of commensurate twisted graphene bilayer. The commensurate moir\'e pattern $(n_1,n_2)$ is defined by the indices $n_1$ and $n_2$ which are coprime numbers \cite{Mele2012}. It has been shown that for each moir\'e pattern $(n_1,n_2)$, there is a twin pattern $(n_1',n_2')$ with greater indices $n_1'$ and $n_2'$ with the same size of the unit cell \cite{Mele2012, Campanera2007}. In the pairs of indices of twin moir\'e patterns,  $(n_1-n_2)/3$ is not integer for the smaller pair of indices, whereas  $(n_1'-n_2')/3$ is integer for the greater indices.
The examples of twin commensurate moir\'e patterns are shown in Fig.~\ref{fig:01}.

If $(n_1-n_2)/3$ is not integer, the unit cell of the commensurate moir\'e pattern $(n_1,n_2)$ is defined by lattice vectors $\vec{L}_1$ and $\vec{L}_2$ [see Fig.~\ref{fig:01}(a)]

\[
   \vec{L}_1 = n_1\vec{a}_1 + n_2\vec{a}_2,\quad
   \vec{L}_2 = -n_2\vec{a}_1 + (n_1+n_2)\vec{a}_2,
\]
where $\vec{a}_1$ and $\vec{a}_2$ are lattice vectors of the bottom graphene layer.

If $(n_1'-n_2')/3$ is integer, the same equations give the vectors one of which corresponds to the diagonal of the moir\'e pattern unit cell [see Fig.~\ref{fig:01}(b)]

\[
   \vec{L}_1' = n_1'\vec{a}_1 + n_2'\vec{a}_2,\quad
   \vec{L}_2' = -n_2'\vec{a}_1 + (n_1'+n_2')\vec{a}_2.
\]

In this case the indices $n_1$ and $n_2$ which determine the moir\'e pattern unit cell
\[
   \vec{L}_1^\text{(u)} = n_1\vec{a}_2 + n_2\vec{a}_1,\quad
   \vec{L}_2^\text{(u)} = -n_2\vec{a}_2 + (n_1+n_2)\vec{a}_1
\]
for given $n_1'$ and $n_2'$ can be found from any of the equations
\[
   \vec{L}_1' = \vec{L}_1^\text{(u)} + \vec{L}_2^\text{(u)},\quad
   \vec{L}_2' = 2\vec{L}_1^\text{(u)} - \vec{L}_2^\text{(u)}
\]
in the form
\[
   n_1 = \frac{{n_1}' + 2{n_2}'}{3},\quad
   n_2 = \frac{n_1' - n_2'}{3}.
\]

The angle $\theta$ of relative rotation of graphene layers of the commensurate moir\'e pattern (that is the angle between the vectors $\vec{a}_1$ and $\vec{a}_1'$) is defined as
\[
   \cos\theta = \frac{n_1^2 + 4n_1n_2 + n_2^2}{2(n_1^2 + n_1n_2 + n_2^2)}.
\]

In the case where $(n_1-n_2)/3$ is not integer, the angle $\varphi$ between the lattice vector $\vec{a}_1$ and the lattice vector of commensurate moir\'e pattern $\vec{L}_1 = n_1\vec{a}_1 + n_2\vec{a}_2$ is

\[
   \varphi = 30^\circ - \frac{\theta}{2}.
\]

The area of the unit cell of moir\'e pattern $(n_1,n_2)$ is
\[
   S = S_gN_c = \frac{\sqrt{3}a^2(n_1^2 + n_1n_2 + n_2^2)}{2R},
\]
where $S_g = \sqrt{3}a^2/2$ is the area of the unit cell of graphene, $a = |\vec{a}_1| = |\vec{a}_2|$ is the graphene lattice constant, $N_c = (n_1^2 + n_1n_2 + n_2^2)/R$ is the number of unit cells of graphene per unit cell of the commensurate moir\'e pattern, the parameter $R = 3$ if $(n_1-n_2)/3$ is integer and $R = 1$ otherwise.

Pairs of twin commensurate moir\'e patterns with the same size of the unit cell and with different symmetry of the stacking just after the relative rotation of layers were considered as different in the original work \cite{Mele2012}. However, these commensurate moir\'e pattern can be obtained one from another by the translational displacement $\vec{t}$ of one of the layers in the layer plane (see Fig.~\ref{fig:01}). Here we study the PES of the interlayer interaction energy as a function of the coordinates describing the in-plane relative displacement of the layers. Evidently, the moir\'e patterns related by the translational displacement $\vec{t}$ correspond to the same PES. Thus, only one of pair moir\'e patterns with the indices $n_1$ and $n_2$, where $(n_1-n_2)/3$ is not integer, are considered here for the PES calculations.

\subsection{Computational details}

\begin{table}
\caption{Calculation details for the considered commensurate moir\'e patterns with coprime indices $(n_1,n_2)$: the angle $\theta$ of relative rotation of graphene layers, the number of atoms $N_c$ in the moir\'e pattern unit cell, simulation cell size in the units of moir\'e pattern unit cells, total number of atoms $N_a$ in the simulation cell and cutoff radius $R_c$ of the Kolmogorov--Crespi potential.}
\renewcommand{\arraystretch}{1.2}
\setlength{\tabcolsep}{8.9pt}
\begin{tabular}{*{6}{c}}
\hline
\hline
$(n_1,n_2)$ & $\theta$ ($^\circ$) & $N_c$ & cell size & $N_a$ & $R_c$ (\AA{}) \\\hline
(2,1) & 21.787 & 7 & $18\times18$ & 9072 & 16 \\
(3,1) & 32.204 & 13 & $18\times18$ & 16848 & 16 \\
(3,2) & 13.174 & 19 & $10\times10$ & 7600 & 16 \\
(5,1) & 42.103 & 31 & $11\times11$ & 15004 & 70 \\
(5,3) & 16.426 & 49 & $9\times9$ & 15876 & 70 \\
(7,2) & 35.567 & 67 & $6\times6$ & 9648 & 50 \\
(7,3) & 26.008 & 79 & $6\times6$ & 11376 & 50 \\
(7,5) & 10.993 & 109 & $5\times5$ & 10900 & 50 \\\hline
 \hline
\end{tabular}
\label{table:vac_struct}
\end{table}

The ratio of the PES corrugations to the average interlayer interaction energy is extremely small for twisted graphene bilayers \cite{Xu2013, Koren2016, Minkin2022}. Simultaneously the size of the simulation cell and the number of computational runs for each considered moir\'e pattern are too high to study the PES of interlayer interaction by {\it ab initio} methods. Thus, we use classical potentials in the present study. At this moment, there are no experimental data on physical properties of systems of twisted layers that can be used to fit parameters of classical potentials for description of the interlayer interaction (or check the adequacy of existing potentials for twisted layers). The parameters of the popular Kolmogorov--Crespi and Lebedeva potentials (Tables~S1 and S2 in Supplemental Material) for the interaction between graphene layers were fitted to the PES of interlayer interaction of coaligned layers (with zero twist angle) obtained by density functional theory (DFT) calculations \cite{Kolmogorov2005, Lebedeva2011}. In the case of the Lebedeva potential,  the experimental data on the frequency of in-plane interlayer vibrations of coaligned layers were also taken into account \cite{Popov2012}. Nevertheless, corrugations of the PES computed for infinite commesurate moir\'e patterns using the Lebedeva potential do not exceed the calculation accuracy \cite{Minkin2021, Minkin2022}. At the same time, they are finite and well-defined for commensurate moir\'e patterns with the smallest unit cells  when the Kolmogorov--Crespi potential is used\cite{Xu2013, Koren2016, Minkin2022}. The explanation of this discrepancy between the Kolmogorov--Crespi and Lebedeva potentials is discussed in Section IIIB. In the present study all calculations are performed using the Kolmogorov--Crespi potential which allows to analyze the shape of the PES determined by the symmetry of commensurate twisted graphene bilayer. However, we emphasize that the results obtained here are only of qualitative nature.

\begin{figure*}
   \centering
 \includegraphics{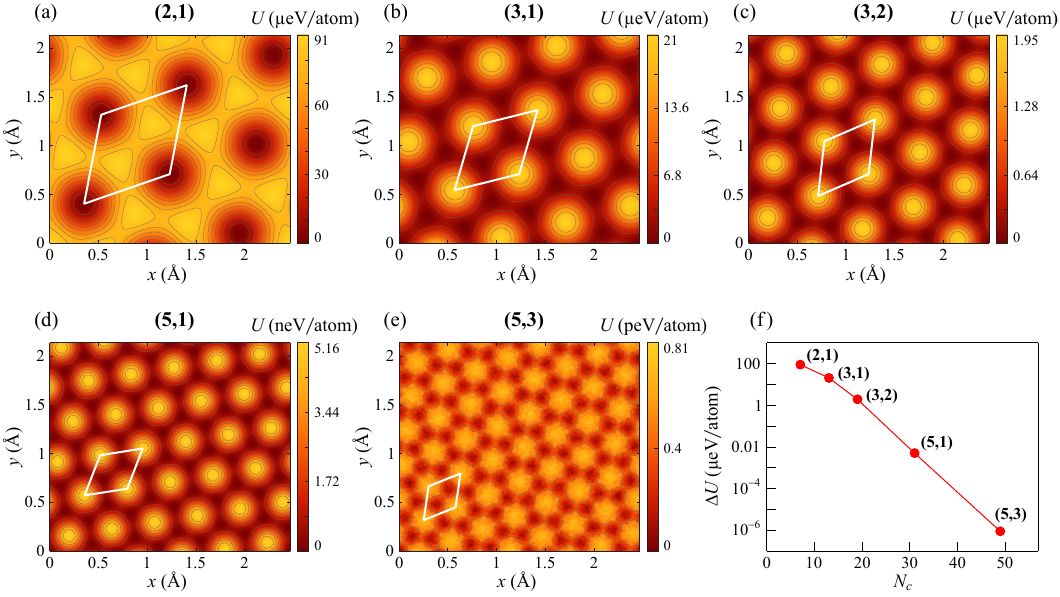}
   \caption{(Color online)  (a-e) Potential energy $U$ (per atom of the upper layer) of interlayer interaction of twisted graphene bilayer as a function of the relative displacement of the layers in the zigzag ($x$, in \AA) and armchair ($y$, in \AA) directions of the lower layer computed at the optimal interlayer distance of 3.46~\AA{} for commensurate moir\'e patterns with coprime indices (a) (2,1), (b) (3,1), (c) (3,2), (d) (5,1) and (e) (5,3). The energy is given relative to the minimum. (f) Amplitude of PES corrugations, $\Delta U$ (per atom of the upper layer), as a function of the number $N_c$ of unit cells of the PES per unit cell of graphene. The indices of the considered moir\'e patterns are indicated.}
   \label{fig:02}
\end{figure*}

The PES calculations have been carried out under the periodic boundary conditions. The simulation cells of height 100 \AA{} have been used for all the considered moir\'e patterns. The bond length between atoms in the graphene layers is taken equal 1.42~\AA{}. The upper graphene layer is placed at the interlayer distance 3.46~\AA{} (which is determined here on the example of (2,1) moir\'e pattern to be optimal for the Kolmogorov--Crespi potential) and is rigidly shifted with respect to the bottom layer with steps of 0.0168~\AA{} and 0.0193~\AA{} in the zigzag and armchair directions of the bottom layer, respectively.  Further calculation details which are different for 8 considered moir\'e patterns with the smallest sizes of unit cell are listed in Table \ref{table:vac_struct}.

\section{Results}

\subsection{PES of twisted graphene bilayer}

The amplitude $\Delta U$ of PES corrugation (i.e. the difference between maximum and minimum values of the interlayer interaction energy, $\Delta U = U_\mathrm{max} - U_\mathrm{min}$) exceeds the calculation accuracy only for 5 out of 8 considered moir\'e patterns with smaller sizes of the unit cells. Calculated PESs for these moir\'e patterns are shown in Fig.~\ref{fig:02}. For 3 out of 8 considered moir\'e patterns with larger sizes of the unit cells (equivalent to the smaller sizes of the unit cells of the PES as shown below), the amplitude of PES corrugations is lower than the artifacts related with the finite value of the cutoff radius of the potential. Calculated PESs for these moir\'e patterns are shown in Fig.~S1 in Supplemental Material.

Two types of the PESs have been found. The PESs of the first type have a triangular lattice of minima and honeycomb lattice of maxima whereas the PESs of the second type, on the contrary, have a triangular lattice of maxima and honeycomb lattice of minima. As discussed in Section IIIB, the PES shape is described for both PES types by the same expression which contains only the first spatial Fourier harmonics and, therefore, only a single energetic parameter. The difference between two types of the PES is determined by the sign of this parameter.

It should be noted that the number of the unit cells of the PES per an unit cell of graphene is the same as the number $N_c$ of unit cells of graphene per the unit cell of the moir\'e pattern (see also Section IIIB). The dependence of the amplitude $\Delta U$ of PES corrugations on the number $N_c$ is shown in Fig.~\ref{fig:02}(f). The amplitude $\Delta U$ decreases nearly exponentially with the decrease of the size of the unit cell of the PES or, equivalently, with the increase of the unit cell of the moir\'e pattern. The analogous exponential decrease of the amplitude of PES corrugations with the increase of the unit cell of the moir\'e pattern was observed previously for rigid finite graphene layers \cite{Xu2013} where the rim contribution to static friction is dominant. Note that extremely low values of the amplitude $\Delta U$ for the moir\'e patterns with the smaller sizes of the PES unit cell make evident the necessity of using classical potentials for the PES calculations.

\subsection{Approximation of PES by the first Fourier harmonics}

PESs of interlayer interaction in diverse hexagonal 2D materials can be closely approximated by the expressions containing only the first spatial Fourier harmonics determined by the system symmetry. The adequacy of such an approximation was demonstrated for coaligned graphene layers \cite{Ershova2010, Lebedeva2011, Popov2012, Zhou2015, Reguzzoni2012}, hexagonal boron nitride (h-BN) \cite{Lebedev2016, Zhou2015}, hydrofluorinated graphene \cite{Lebedev2020}, graphene/h-BN heterostructure \cite{Jung2015, Kumar2015, Lebedev2017} and double-layer graphene with krypton spacer \cite{Popov13b}.

\begin{figure}
   \centering
 \includegraphics{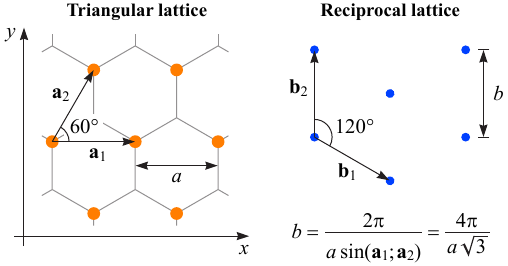}
   \caption{Triangular lattice (orange dots) and its reciprocal lattice (blue dots). The lattice vectors $\vec{a}_1$ and $\vec{a}_2$ ($|\vec{a}_1| = |\vec{a}_2| = a$) as well as reciprocal lattice vectors  $\vec{b}_1$ and $\vec{b}_2$ are shown. Graphene honeycomb lattice is shown by grey lines. Coordinate $x$ corresponds to zigzag direction of graphene layer.}
   \label{fig:03}

\end{figure}

These approximations are based on the following considerations. The translational symmetry of the PES for an atom adsorbed on a triangular lattice is the same as of the triangular lattice, that is $U_a(\vec{r}) = U_a(\vec{r}  + n_1\vec{a}_1 + n_2\vec{a}_1)$, where $\vec{a}_1$ and $\vec{a}_2$ are the lattice vectors ($|\vec{a}_1| = |\vec{a}_2| = a$ and the angle between the vectors is 60$^\circ$, Fig.~\ref{fig:03}), for any integer $n_1$ and $n_2$. This means that the Fourier transform of $U_a(\vec{r})$ consists of harmonics corresponding to vertices of the reciprocal lattice with the lattice vectors $\vec{b}_1$ and $\vec{b}_2$ such that $\vec{a}_i \cdot \vec{b}_j = 2\pi\delta_{ij}$ ($|\vec{b}_1| = |\vec{b}_2| = 4\pi/\sqrt{3}a$ and the angle between these vectors is 120$^\circ$, Fig.~\ref{fig:03}). Taking into account only the first Fourier harmonics with wavevectors
 $\vec{b}_1$, $\vec{b}_2$ and  $\vec{b}_1 + \vec{b}_2$, the PES for an atom on a triangular lattice can be approximated as
\begin{equation} \label{eq_at0}
   \begin{split}
\delta U_a(\vec{r})= U_{a,1}\mathrm{Re}\bigg[e^{i\vec{b}_1\vec{r}}+e^{i\vec{b}_2\vec{r}}+e^{i(\vec{b}_1+\vec{b}_2)\vec{r}}\bigg],
   \end{split}
\end{equation}
where $\delta U_a$ is the deviation from the average interaction energy between the atom and lattice and point $\vec{r} = 0$ corresponds to the case when the atom is located on top of one of the lattice atoms.

For $x$ and $y$ axes chosen along one of the lattice vectors ($\vec{a}_1$) and in the perpendicular direction \cite{Verhoeven2004},
\begin{equation} \label{eq_at}
   \begin{split}
\delta U_a(x, y)= U_{a,1}\Big[2\cos{(k_{y} y)}\cos{(k_{x} x)} +\cos{(2k_{y} y)}\Big],
   \end{split}
\end{equation}
 where $k_{x} = 2\pi/a$ and $k_{y} = 2\pi/\sqrt{3}a$.

To get the potential energy of an atom on a honeycomb lattice, it is needed to sum up the expressions for two sublattices separated by $a/\sqrt{3}$ along the $y$ axis (in the armchair direction). This leads to a sign change for the second term of Eq.~(\ref{eq_at}). For two coaligned honeycomb layers, it is needed to sum up once more the contributions of two sublattices of the adsorbed layer. This leads to another sign change in the equation. Thus, the PES of coaligned honeycomb layers is described the equation similar to Eq.~(\ref{eq_at}) (Refs.~\onlinecite{Ershova2010, Lebedeva2011, Popov2012}).

Now let us consider twisted honeycomb lattices. Let $\vec{a}_1$ and $\vec{a}_2$ be the lattice vectors of the bottom layer and $\vec{a}'_1$ and $\vec{a}'_2$ of the upper one. The PES of twisted honeycomb lattices is periodic with respect to translation along any of these lattice vectors: $U(\vec{r}) = U(\vec{r}  + n_1\vec{a}_1 + n_2\vec{a}_2) = U(\vec{r}  + n'_1\vec{a}'_1 + n'_2\vec{a}'_2)$,
where $n_1$, $n_2$, $n'_1$ and $n'_2$ are any integer numbers. Thus, the harmonics that contribute to the Fourier transform if the PES of twisted honeycomb lattices should comply with the condition $\vec{G} = m_1\vec{b}_1 + m_2\vec{b}_2 = m'_1\vec{b}'_1 + m'_2\vec{b}'_2$, where $\vec{b}_1$ and $\vec{b}_2$ are the lattice vectors of the reciprocal lattice of the bottom layer, $\vec{b}'_1$ and $\vec{b}'_2$ of the upper one and $m_1$, $m_2$, $m'_1$ and $m'_2$ are some integer numbers. This means these Fourier harmonics correspond to overlapping vertices of the reciprocal lattices of the twisted layers. The reciprocal lattices of the twisted honeycomb layers are also two twisted honeycomb lattices forming a commensurate moir\'e pattern similar to one in real space. Therefore, the first Fourier terms contributing to the PES in this case correspond to the lattice vectors of this moir\'e pattern of the reciprocal lattices. They have the length $B = bL/a$, where $L$ is the period of the moir\'e pattern, and are rotated with respect to the reciprocal lattice vectors by the same angle $\varphi$ as the moir\'e pattern vectors are rotated with respect to the lattice vectors in real space. As a result, the PES of twisted honeycomb layers can be approximated in the same form as Eq.~(\ref{eq_at}) if we consider $x'$ and $y'$ axes aligned along one of the moir\'e pattern vectors and in the perpendicular direction as well as increase the wavevectors by the factor of $L/a = \sqrt{N_c}$:
\begin{equation} \label{eq_approx}
\delta U(x',y') = U_1\bigg(2\cos{(k'_yy')}\cos{(k'_xx')} +\cos{(2k'_yy')}\bigg),
\end{equation}
where $\delta U = U - U_\text{av}$ is the deviation from the average PES energy $U_\text{av} ={}$const, $x' = x \cos\varphi - y \sin\varphi$,  $y' = y \cos\varphi + x \sin\varphi$, $k'_x = \sqrt{N_c}k_x$ and $k'_y = \sqrt{N_c}k_y$. Thus the PES of interlayer interaction of an infinite graphene bilayer with a commensurate moir\'e pattern has the same shape as the PES of for graphene bilayer with coaligned layers (presented in Ref.~\onlinecite{Ershova2010, Lebedeva2011, Popov2012, Reguzzoni2012}) and differs only by the PES amplitude and the period which is lower by the factor of $\sqrt{N_c}$.

The derived Eq.~(\ref{eq_approx}) is used here for the approximation of the calculated PESs. To fit the amplitude $\Delta U$ of PES corrugations, the \emph{single} parameter of the approximation is chosen as $U_1 = 2\Delta U/9$. The relative deviation $\epsilon$ of the approximated and computed PESs is found as the root-mean-square deviation divided by $\Delta U$. As can be seen in Table \ref{table:prop}, this relative deviation varies from 0.02\% to the maximum of 1\% for all the considered moir\'e patterns. The relative deviations in previous studies of PESs of coaligned layers are about 1\% for graphene \cite{Popov2012, Lebedeva2011}, 0.1--0.3\% for h-BN \cite{Lebedev2016}, 0.3\% for graphene/h-BN heterostructure \cite{Lebedev2017} and 3\% for hydrofluorinated graphene \cite{Lebedev2020}. We believe that the simple shape of the PES obtained here is a universal property of commensurate twisted bilayers consisting of various 2D materials analogous to that of coaligned commensurate bilayers.

It should be noted that for coaligned graphene layers, $U_{a,1}$ in Eq.~(\ref{eq_at}) is positive and corresponds to the repulsion between the atoms of the upper and lower layers.\cite{Ershova2010, Lebedeva2011, Popov2012, Zhou2015, Reguzzoni2012} In the case of a moir\'e pattern, the PES is determined by the sum of contributions from many atoms within the moir\'e pattern unit cell and $U_1$ in Eq.~(\ref{eq_approx}) can be positive or negative (Fig.~\ref{fig:02} and Fig.~S1 in Supplemental Material) depending in which symmetry points there is more repulsion between the layers.

Let us discuss the discrepancy of the PESs obtained here using the Kolmogorov--Crespi potential and the results of the calculations using the Lebedeva potential, where no corrugations which exceed the calculation accuracy are observed for the (2,1) moir\'e pattern \cite{Minkin2021, Minkin2022}. The PES of interlayer interaction for coaligned graphene layers obtained by the DFT-D calculations is excellently approximated by the first Fourier harmonics \cite{Ershova2010, Lebedeva2011, Popov2012, Reguzzoni2012}. The Lebedeva potential was specifically designed to reproduce this property of the PES so that the relative root-mean-square deviation of the approximated and computed PESs is within several percents for the Lebedeva potential. A similar deviation for the Kolmogorov--Crespi potential is 20 times greater \cite{Lebedeva2011}. The greater deviation of the Kolmogorov--Crespi potential is related with considerably larger amplitudes of other Fourier harmonics except the first one (which exactly reproduces the approximated PES) including those compatible with the symmetry of commensurate twisted graphene bilayers and, therefore, responsible for the shape of corresponding PESs. This explains the discrepancy of the results obtained using the Kolmogorov--Crespi and Lebedeva potentials.

Eqs.~(\ref{eq_at}) and (\ref{eq_approx}) are derived based on the system symmetry and make sense not only for the interlayer interaction energy but also for other properties of 2D materials. For example, Eq.~(\ref{eq_at}) was used to approximate the interlayer tunneling contribution to the Hamiltonian of graphene layers in Ref.~\onlinecite{Jung2014}. Considering the Brillouin zone for twisted layers, the authors derived for them the explicit Hamiltonian and analyzed their band structures. A similar study was also performed for graphene/h-BN heterostructure \cite{Jung2014, Jung2015}. The use of Eq.~(\ref{eq_approx}) taking into account the symmetry of commensurate twisted bilayers might be used to further simplify such models.

\subsection{Properties related with PES}

A number of physical properties associated with relative in-plane motion of the layers are determined by the PES at a constant interlayer distance \cite{Popov2012, Lebedev2016, Lebedev2017, Lebedev2020}. Since the PES is described by a simple expression involving just one energetic parameter [see Eq.~(\ref{eq_approx})], all these properties
can be described analytically as functions of this parameter. Below we use the PES for the considered moir\'e patterns to estimate the shear mode frequency, shear modulus, shear strength and barrier for relative rotation of the commensurate twisted layers to an incommensurate state.

\begin{table*}
\caption{Approximation parameters $U_1$, relative root-mean-square deviations $\epsilon$, shear mode frequencies $f$, shear moduli $C_{44}$, shear strengths $\tau$ and barriers $\Delta U_\mathrm{rot}$ for relative rotation of the commensurate twisted layers to an incommensurate state estimated for different moir\'e patterns based on calculations with the Kolmogorov--Crespi potential.}
\renewcommand{\arraystretch}{1.2}
\setlength{\tabcolsep}{12pt}
\begin{tabular}{*{7}{c}}
\hline
\hline
moir\'e pattern & $U_1$ (eV/atom\footnote{per atom of the upper layer}) & $\epsilon$ & $f$ (cm$^{-1}$) & $C_{44}$ (Pa) & $\tau$ (Pa) & $\Delta U_\mathrm{rot}$ (eV/atom$^\mathrm{a}$)  \\
\hline
(2,1) &  $-$2.02$\cdot10^{-5}$ & 1.04$\cdot10^{-2}$ & 9.153  & 3.91$\cdot10^{8}$  & 1.67$\cdot10^{7}$ & 6.07$\cdot10^{-5}$ \\
(3,1) &  4.66$\cdot10^{-6}$  & 5.34$\cdot10^{-3}$ &  4.233 &  8.37$\cdot10^{7}$ & 1.12$\cdot10^{6}$& 6.99$\cdot10^{-6}$ \\
(3,2) &  4.34$\cdot10^{-7}$  & 3.33$\cdot10^{-3}$ &  1.561  & 1.14$\cdot10^{7}$  & 1.26$\cdot10^{5}$ & 6.50$\cdot10^{-7}$ \\
(5,1) & 1.15$\cdot10^{-9}$   & 2.04$\cdot10^{-3}$ & 0.103 & 4.91$\cdot10^{4}$ & 4.25$\cdot10^{2}$ &  1.72$\cdot10^{-9}$ \\
\hline
\hline
\end{tabular}
\label{table:prop}
\end{table*}

The frequency of the shear mode $E_{2g}$, in which adjacent layers slide rigidly in the opposite in-plane directions, can be found from the PES curvature in a given energy minimum \cite{Popov2012, Lebedev2016, Lebedev2017, Lebedev2020} as
\begin{equation} \label{eq_freq}
   \begin{split}
      f = \frac{1}{2\pi}\sqrt{\frac{1}{\mu}\frac{\partial^2 U}{\partial x'^2}} =
      \frac{1}{a}\sqrt{\frac{1}{\mu}U_{\mathrm{eff}}},
   \end{split}
\end{equation}
where $U_{\mathrm{eff}} = (a/2\pi)^2 \partial^2 U/\partial x'^2$ is the second-order derivative of the energy per carbon atom of the upper layer in energy units and $\mu$ is the reduced mass. The latter for bilayer graphene is as $\mu = m_\mathrm{C}/2$, where $m_\mathrm{C}$ is the mass of a carbon atom.

From Eq.~(\ref{eq_approx}), it follows that the PES curvature corresponds to $U_{\mathrm{eff}} = N_c U_1 = (2/9) N_c \Delta U$ for moir\'e patterns with $U_1 > 0$ and $-2N_c U_1$ for $U_1 < 0$. The shear mode frequencies estimated for different moir\'e patterns using the values of the parameter $U_1$ derived from the calculations with the Kolmogorov--Crespi potential are listed in Table \ref{table:prop}. They all are within 10~cm$^{-1}$ and are considerably smaller than the shear mode frequency for the coaligned graphene layers: 35~cm$^{-1}$ (Ref.~\onlinecite{Lebedeva2011, Lebedeva2012}) and 21--34~cm$^{-1}$ (Ref.~\onlinecite{Lebedeva2016a}) according to DFT calculations and $28\pm3$ cm$^{-1}$ (Ref.~\onlinecite{Boschetto2013}) and 32~cm$^{-1}$ (Ref.~\onlinecite{Tan2012}) according to the experiments. Indeed, as follows from Eq.~(\ref{eq_freq}), the frequency depends on the square root $\sqrt{\Delta UN_c}$ of the product of the amplitude $\Delta U$ of PES corrugations and the number $N_c$ of unit cells of graphene per the unit cell of the commensurate moir\'e pattern. Since $\Delta U$ decreases exponentially with the growth of $N_c$ [Fig.~\ref{fig:02}(f)], the shear mode frequency also gets reduced upon increasing $N_c$.

The PES curvature also determines the shear modulus \cite{Lebedev2016, Lebedev2020}
\begin{equation} \label{eq_C44}
   \begin{split}
      C_{44} = \frac{d}{\sigma}\frac{\partial^2 U}{\partial x'^2} =
      \frac{16\pi^2 d}{\sqrt{3}a^4} U_{\mathrm{eff}},
   \end{split}
\end{equation}
where $\sigma =\sqrt{3}a^2/4$ is the area per carbon atom and $d = 3.46$~\AA~is the interlayer distance. The shear moduli estimated for different moir\'e patterns do not exceed 0.4 GPa (Table \ref{table:prop}). This is an order of magnitude smaller than the DFT result for the coaligned graphene bilayer of 3.8--4.1 GPa (Ref.~\onlinecite{Lebedeva2016a}) because the shear modulus is proportional to $\Delta UN_c$.

The PES also determines the static friction force $f_s$ for moving the layers as a whole, i.e. the maximal first derivative of the potential energy along the minimum energy path (MEP) between adjacent energy minima and, correspondingly, the shear strength $\tau$ related to it as  $\tau =f_s/\sigma$ (here the force is taken per atom of the upper layer). Analogous estimates of shear strength have been performed first for commensurate double-walled nanotubes \cite{Bichoutskaia2009a}. For $U_1 > 0$, the MEP between adjacent energy minima corresponds to the line $x'=0$ and $y'$ from $a/\sqrt{3N_c}$ to $2a/\sqrt{3N_c}$ in Eq.~(\ref{eq_approx}). The force along the MEP is given by
\begin{equation} \label{eq_strength_pos}
      -\frac{\partial U}{\partial y'}\Big|_{x'=0} = 2k'_y U_1\bigg(\sin{(k'_yy')} +\sin{(2k'_yy')}\bigg)
\end{equation}
and the force extrema are determined by the equation $\cos{(k'_yy')} +2\cos{(2k'_yy')}=0$. The latter equation gives that the maximal absolute force is achieved for $\cos{(k'_yy')} = -(1 + \sqrt{33})/8$ for the considered MEP. From this, we find that the shear strength is $\tau =6.183\sqrt{N_c} U_1/a^3$ for $U_1 > 0$.

For $U_1 < 0$, the MEP corresponds to $y'=0$ and $x'$ from $0$ to $a/\sqrt{N_c}$ in Eq.~(\ref{eq_approx}). In this case,
\begin{equation} \label{eq_strength_neg}
      -\frac{\partial U}{\partial x'}\Big|_{y'=0} = 2k'_x U_1\sin{(k'_xx')}.
\end{equation}
The maximal absolute force is achieved for $x' = a/4$. Correspondingly, the shear strength is  $\tau =16\pi \sqrt{N_c} U_1/\sqrt{3}a^3$.

The shear strength $\tau$  values estimated for different moir\'e patterns are within 0.02 GPa (Table \ref{table:prop}). From typical DFT values for the amplitude of PES corrugations for coaligned graphene layers of about 15 meV per atom of the upper layer \cite{Popov2012, Lebedeva2016a}, we deduce that the shear strength in that case should be about 0.22 GPa, i.e. an order of magnitude greater. As seen from the above equations, the shear strength for moir\'e patterns is proportional to $\sqrt{N_c}\Delta U$.

When the graphene layers are rotated with respect to each other by an arbitrary angle that does not correspond to any commensurate moir\'e pattern, the area contribution to the PES vanishes and the PES of an infinite incommensurate twisted bilayer becomes flat. Therefore, the interaction energy in such a fully incommensurate state can be found as an average over the PES: $U_{\mathrm{rot}} = \langle U \rangle_{x,y}$ (Refs.~\onlinecite{Popov2012, Lebedev2016, Lebedev2017, Lebedev2020}). The barrier $\Delta U_{\mathrm{rot}}$ for relative rotation of the layers to a fully incommensurate state can, thus, be obtained by subtracting the energy in the minimum from $U_{\mathrm{rot}}$. From  Eq.~(\ref{eq_approx}), one gets $\Delta U_{\mathrm{rot}} = 1.5U_1$ for $U_1 > 0$ and  $-3U_1$ for $U_1 < 0$. The values of the barrier estimated for the moir\'e patterns considered are within 0.06 meV per atom of the upper layer (Table \ref{table:prop}). Obviously they are much smaller than the previous predictions for the coaligned graphene bilayer of 4 meV/atom (Refs.~\onlinecite{Lebedeva2010, Lebedeva2010a}) and 5~meV/atom (Ref.~\onlinecite{Popov2012}).

Structural superlubricity can be lost via rotation of the layers with the same lattice constant to the commensurate ground state with coaligned layers \cite{Hirano1990, Verhoeven2004, Dienwiebel2005, Filippov2008, Bonelli2009, Guo2007, Shibuta2011, Xu2013, Wang2019, Feng2013}. The robust superlubricity has been recently achieved for systems with a lattice mismatch such as heterostructures composed of layers of different 2D materials \cite{Song2018} or layers of the same 2D material under different tension applied \cite{Wang2019, Androulidakisl2020}. For such robust superlubric systems, the relative rotation of the layers to a commensurate interface with the loss of superlubricity is not possible. Here we propose that the robust superlubricity can be also achieved for systems in which rotation of the layers to the commensurate ground state with coaligned layers is possible but hindered by a barrier. For twisted commensurate layers, such a rotation should occur through a fully incommensurate state. Although the barriers between a local minimum of twisted commensurate state and a fully incommensurate state calculated here are rather small, they might result sufficient to prevent the relative rotation of layers for a sufficiently large contact area and thus can ensure the macroscopic robust superlubricity. Further macroscale investigations would be needed to confirm this hypothesis.

It should be kept in mind that the values of the physical quantities obtained in the present Section and listed in Table \ref{table:prop} are based on the calculations with the Kolmogorov-Crespi potential \cite{Kolmogorov2005}, while its adequacy for twisted graphene layers has not been proven. Nevertheless,  the potential gives a reasonable dependence of the amplitude $\Delta U$ of PES corrugations on the size of the unit cell of the moir\'e pattern [Fig.~\ref{fig:02}(f)] and thus should properly describe the trend in the evaluated physical quanities for different moir\'e patterns. Once a more reliable potential for twisted graphene layers is available, it can be used to obtain more accurate estimates based on the equations given above. On the other hand, as soon as any of these physical quantities is accessed experimentally (by analogy with the measurements for coaligned graphene layers \cite{Boschetto2013, Tan2012}), classical potentials can be refined to improve the description of the PES of the twisted layers on the basis of the above formalism. We also believe that the simple shape of the PES can be a universal property for commensurate twisted bilayers consisting of diverse 2D materials. This means that physical properties of moir\'e patterns of other 2D materials can be estimated in a similar way.

The results of the geometrical analysis persented here are derived without taking into account structural relaxation. Our calculations show that the account of structural relaxation does not lead to changes in the shape of PES and it is still described by the first Fourier harmonics (whereas some increase of the amplitude $\Delta U$ of PES corrugations occurs). Thus, all the equations presented here are still valid. The influence of the structural relaxation on structural and tribological properties of commensurate moir\'e patterns will be considered elsewhere.

\section{Discussion and conclusions}

PESs of interlayer interaction have been calculated for a set of commensurate moir\'e patterns of twisted graphene bilayer using the registry-dependent Kolmogorov--Crespi potential. The amplitude of PES corrugations is found to exceed the calculation accuracy only for 5
moir\'e patterns with the smaller unit cell sizes. All calculated PESs have the same simple shape which corresponds to the symmetry of commensurate moir\'e patterns and with the size of the unit cell of PES which is inversely related to the unit cell size of the moir\'e pattern. The amplitude of PES corrugations exponentially decreases with increasing the unit cell size of the moir\'e pattern. An analytical expression which is based on the first Fourier harmonics describing the PES has been derived. The calculated PESs can be approximated by the derived expression with the accuracy within 1\% relative to the amplitude of PES corrugations. Since the derived expression contains a single energetic parameter, it has been used to estimate a set of physical quantities determined by the PES such as shear mode frequency, shear modulus, shear strength and barrier for relative rotation of the commensurate twisted layers to a fully incommensurate state. We propose that the latter barrier might prevent the rotation of the layers from the twisted commensurate state to the ground commensurate state through a fully incommensurate state and, therefore, can possibly lead to the macroscopic robust superlubricity for a sufficiently large contact area.

The approximation by the first Fourier harmonics can be applied not only for consideration of the interlayer interaction energy. For example, such an approximation for coaligned layers was used for the anaysis of electronic properties of twisted graphene \cite{Jung2014} and graphene/h-BN heterostructure \cite{Jung2014, Jung2015}. Moreover, since the approximation of the PES of interlayer interaction by the first Fourier harmonics is a universal property for coaligned layers of diverse 2D materials \cite{Lebedev2020}, we believe that the simple shape of the PES obtained here for twisted commensurate graphene bilayer can be also universal for any commensurate moir\'e patterns consisting of layers of diverse 2D materials.

The raw data on calculated PES required to reproduce our findings are available to download from Ref.~\onlinecite{Minkin23Mendeley}.

\hfil
\section*{Acknowledgments}

A.S.M., A.M.P. and Y.E.L. acknowledges the support by the Russian Science Foundation grant No.~23-42-10010, https://rscf.ru/en/project/23-42-10010/, for the results described in subsections IIIA ``PES of twisted graphene bilayer'' and IIIB ``Approximation of PES by the first Fourier harmonics''. I.V.L. acknowledges Bikaintek grant ``Transport'' from the Basque Government. A.M.P. and Y.E.L. acknowledges the support by project FFUU-2021-0003 of the Institute of Spectroscopy of the Russian Academy of sciences for the results described in subsection IIIC ``Properties related with PES''. S.A.V. and N.A.P. acknowledge support by the Belarusian Republican Foundation for Fundamental Research (Grant No. F23RNF-049) and by the Belarusian National Research Program ``Convergence-2025''. This work has been particularly carried out using computing resources of the federal collective usage center Complex for Simulation and Data Processing for Mega-science Facilities at NRC ``Kurchatov Institute'', http://ckp.nrcki.ru.

The authors declare no conflict of interest.

\bibliography{prb2021-popov}
\end{document}


\title{Supplemental Material for ``Interlayer interaction, shear vibrational mode, and tribological properties of two-dimensional bilayers with a commensurate moir\'e pattern''}

\author{Alexander S. Minkin}
\email{amink@mail.ru}
\affiliation{Keldysh Institute of Applied Mathematics of Russian Academy of Sciences,
4 Miusskaya sq., Moscow, 125047, Russia}

\author{Irina V. Lebedeva}
\email{liv\_ira@hotmail.com}
\affiliation{Simune Atomistics, Avenida de Tolosa 76, San Sebastian 20018, Spain}

\author{Andrey M. Popov}
\email{popov-isan@mail.ru}
\affiliation{Institute for Spectroscopy of Russian Academy of Sciences, Troitsk, Moscow 108840, Russia}

\author{Sergey A. Vyrko}
\email{vyrko@bsu.by}
\affiliation{Physics Department, Belarusian State University, Nezavisimosti Ave.~4, Minsk 220030, Belarus}

\author{Nikolai A. Poklonski}
\email{poklonski@bsu.by}
\affiliation{Physics Department, Belarusian State University, Nezavisimosti Ave.~4, Minsk 220030, Belarus}

\author{Yurii E. Lozovik}
\email{lozovik@isan.troitsk.ru}
\affiliation{Institute for Spectroscopy of Russian Academy of Sciences, Troitsk, Moscow 108840, Russia}
\affiliation{Moscow Institute of Electronics and Mathematics, National Research University Higher School of Economics, Bol.~Trekhsvjatitel'skij per., 1-3/12, build.~8, Moscow, 101000, Russia}

\maketitle

\onecolumngrid
\renewcommand{\thepage}{S\arabic{page}}
\renewcommand{\thefigure}{S\arabic{figure}}
\renewcommand{\thetable}{S\arabic{table}}
\renewcommand{\theequation}{S\arabic{equation}}

\vspace{-20pt}
\section*{Kolmogorov--Crespi potential}
To take into account that the $\pi$-overlap between
graphene layers is anisotropic, the Kolmogorov--Crespi
potential\cite{Kolmogorov2000, Kolmogorov2005} depends differently on in-plane and out-of-plane directions. It is assumed that the interaction energy of atoms at distance $r$, transverse separation $\rho$ and interlayer spacing $z$ ($r^2 = \rho^2 + z^2$) can be written as
\begin{equation} \label{eq_kc}
U = - A \left(\frac{z_0}{r}\right)^6  + \left(C+2\exp\left(-\left(\frac{\rho}{\delta}\right)^2\right) \sum_{n=0}^2{C_{2n}\left(\frac{\rho}{\delta}\right)^{2n}}\right)\exp\left(-\lambda(r-z_0)\right).
\end{equation}
The parameters of the potential  fitted to graphite compressibility, equilibrium interlayer distance and interlayer binding energy as well as the data on relative sliding of graphene layers obtained by density functional theory (DFT) calculations in the local-density approximation \cite{Kolmogorov2005} are listed in \textbf{Table S1}.

\begin{table}[h]
\caption{Parameters of the Kolmogorov--Crespi potential.}
\renewcommand{\arraystretch}{1.2}
\setlength{\tabcolsep}{6pt}
\begin{tabular}{*{8}{c}}
\hline
\hline
 $A$ (meV) & $z_0$ (\AA) & $C$ (meV) & $C_0$ (meV) & $C_2$ (meV) & $C_4$ (meV) & $\delta$ (\AA) & $\lambda$ (\AA$^{-1}$)\\\hline
10.238 & 3.34 & 3.030 & 15.71 & 12.29 & 4.933 & 0.578 & 3.629 \\\hline
 \hline
\end{tabular}
\label{table:kc}
\end{table}

\vspace{-10pt}
\section*{Lebedeva potential}
In the Lebedeva potential\cite{Lebedeva2011, Popov2012}, the interaction energy of atoms at distance $r$, transverse separation $\rho$ and interlayer spacing $z$ is given by
\begin{equation} \label{eq_leb}
U = A \left(\frac{z_0}{r}\right)^6 + B\exp\left(-\alpha(r-z_0)\right) + C\left(1+D_1\rho^2+D_2\rho^4\right)\exp\left(-\lambda_1\rho^2\right) \exp\left(-\lambda_2(z^2 - z_0^2)\right).
\end{equation}
The parameters of the potential were fitted to the experimental data on graphite compressibility, equilibrium interlayer distance and interlayer binding energy and the data on relative sliding of graphene layers obtained by DFT calculations with the dispersion correction \cite{Lebedeva2011}.  Later the potential was slightly adjusted to take into account the experimental data on shear mode frequencies in few-layer graphene and graphite \cite{Popov2012}. Compared to the Kolmogorov--Crespi potential, the Lebedeva potential better reproduces the shape of the potential energy surface for coaligned graphene layers and its dependence on the interlayer distance following from the DFT calculations\cite{Lebedeva2011}. The parameters of the Lebedeva potential are listed in  \textbf{Table S2}.

\begin{table}[h]
\caption{Parameters of the Lebedeva potential.}
\renewcommand{\arraystretch}{1.2}
\setlength{\tabcolsep}{6pt}
\begin{tabular}{*{9}{c}}
\hline
\hline
 $A$ (meV) & $B$ (meV) & $z_0$ (\AA) & $\alpha$ (\AA$^{-1}$) & $C$ (meV) & $D_1$ (\AA$^{-2}$) & $D_2$ (\AA$^{-4}$) & $\lambda_1$ (\AA$^{-2}$) &$\lambda_2$ (\AA$^{-2}$)\\\hline
$-$10.510 & 11.652 & 3.34 & 4.16 & 29.5 & $-$0.86232 & 0.10049 & 0.48703 & 0.46445 \\\hline
 \hline
\end{tabular}
\label{table:leb}
\end{table}

\newpage

\vspace{-10pt}
\section*{Potential energy surfaces}
Potential energy surfaces (PESs) for the moir\'e patterns with the amplitude of the PES corrugations which is lower than the artifacts related with the finite value of the cutoff radius of the potential used for the calculations are shown in \textbf{Figure S1}.
\vspace{6pt}

\begin{figure*}[!h]
   \centering
\includegraphics{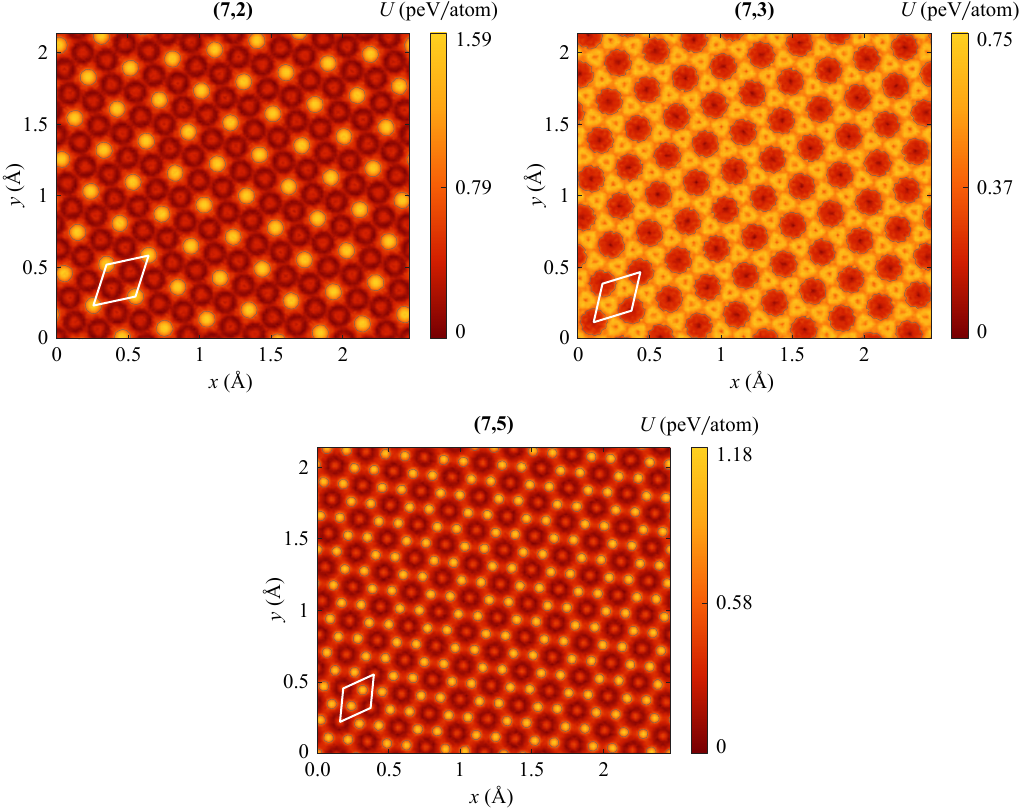}
\vspace{-6pt}
   \caption{Potential energy $U$ (in picoelectronvolts per atom of the upper layer) of interlayer interaction of twisted graphene bilayer as a function of the relative displacement of the layers in the zigzag ($x$, in angstroms) and armchair ($y$, in angstroms) directions of the lower layer computed at the optimal interlayer distance of 3.46~\AA{} for moir\'e patterns with coprime indices (7,2), (7,3), and (7,5). The energy is given relative to the minimum. The elementary cells of the PESs are shown by white diamonds.}
   \label{fig:s1}
\vspace{-20pt}
\end{figure*}

\bibliography{prb2021-popov}